\documentclass[]{article}
\usepackage{latexsym}
\usepackage{epsfig}

\begin{document}
\baselineskip=5mm
\quad


\vspace{8pt}

\centerline{\Large \bf Estimations of
$\overline{\Omega}^+/\Omega^-$ at RHIC from a QGP Model with
Diquarks}

\vspace{24pt} \centerline{ \large Hong Miao$^1$, Chongshou
Gao$^{1, 2}$}

\vspace{16pt}

 \centerline{ $^1$School of Physics, Peking
University, Beijing 100871, China}
 \centerline{ $^2$Institute of Theoretical Physics,
Academia Sinica, Beijing 100080, China}

\vspace{16pt}
\begin{abstract}

Assuming that axial-vector and scalar diquarks exist in the
Quark-Gluon Plasma near the critical temporature $T_c$, baryons
can be produced from quark-diquark interactions. In RHIC
conditions ($\sqrt{s_{NN}} = 130GeV$ and $200GeV$), the ratio
$\overline{\Omega}^+/\Omega^-$ may be larger than 1, based on the
concept that QGP with diquarks would exist. This unusual result
might be a helpful evidence for QGP existing in RHIC.
 \vspace*{8pt}
\\ {\bf PACS number(s)\/}: 12.38.Mh 25.75.-q
\\{\bf Key words\/}: QGP, plasma, diquark, strangeness, baryon
\end{abstract}

\vspace{12pt}

\leftline{ \bf 1. Introduction}

\vspace{8pt}

Diquarks\cite{sign:sign diquark_form} may exist as bound
states\cite{sign:sign bound_state}\cite{sign:sign screen} in the
Quark-Gluon Plasma (QGP) as well as quarks and gluons. If
axial-vector and scalar diquarks exist near the critical
temperature $T_c$ and approximate thermal equilibrium could form,
baryon production can be described as the process of quark and
diquark forming $(\frac{1}{2})^+$ and $(\frac{3}{2})^+$ baryon
states\cite{sign:sign diquark_baryon}. Ratios of different
baryons can be estimated through this method. Since strange
baryon production is widely discussed\cite{sign:sign
Rafelski_phL}\cite{sign:sign Rafelski_Acta1}\cite{sign:sign
Rafelski_Acta2}\cite{sign:sign Rafelski_Acta3}\cite{sign:sign
Strange}\cite{sign:sign RatiosModel} and has a upper limit in the
Hadronic Gas Model\cite{sign:sign Strange}, the ratio beyond that
limit can support the idea of QGP productions in the relativistic
heavy ion collisions.

QGP with diquarks has a much higher energy density than general
Quark-Gluon Plasma\cite{sign:sign diquark_density}. Another
interesting phenomena there is about the ratio of
$\overline{\Omega}^+/\Omega^-$. In the conditions of hadronic
matter, the ratio will smaller than 1\cite{sign:sign
Omega/OmegaSPS}. General QGP model will also predict the ratio
$\overline{\Omega}^+/\Omega^- = 1$ \cite{sign:sign
Rafelski_phL}\cite{sign:sign Rafelski_Acta2} or near it.

While, since there are diquarks together with quarks in the QGP,
strange particles are not only $s$ and $\bar{s}$, but also
$V(ss)$, $\overline{V}(ss)$, $V(us)$, $\overline{V}(us)$,
$V(ds)$, $\overline{V}(ds)$, $S(us)$, $\overline{S}(us)$, $S(ds)$
and $\overline{S}(ds)$. So strangeness conservation will not
simply requires $N(s) = N(\bar{s})$. When $\mu_B > 0$, $(us)$ and
$(ds)$ diquarks' amounts will be larger than those of
$(\bar{u}\bar{s})$ and $(\bar{d}\bar{s})$ diquarks, but $s$
quark's and $(ss)$ diquark's amounts will be smaller than those of
$\bar{s}$ quark and $(\bar{s}\bar{s})$ diquark. That means the
amount of $\Omega^-(sss)$ would be a little smaller than
$\overline{\Omega}^+(\bar{s}\bar{s}\bar{s})$. This is a general
subsequence based on diquark's appearance. Some models from
strings\cite{sign:sign Omega/OmegaString} also predict
$\overline{\Omega}^+/\Omega^- > 1$ in p-p collisions, when
diquarks are imported.

When concerning that the baryon chemical potential and
strangeness chemical potential may have changed during the
process of freeze-out, the result will be very complicated. And as
a simplification, such influence has been neglected in the
calculations. In addition, hadron interactions after phase
transition may lower the ratios of
$\overline{\Omega}^+/\Omega^-$, and can not be easily estimated
so far.

\vspace{12pt}

\leftline{ \bf 2. Diquark Model and Baryon Production in QGP}

\vspace{8pt}

In the SU(6) quark-diquark model, baryon wave functions can be
described as combinations of quarks and diquarks\cite{sign:sign
diquark_form}\cite{sign:sign diquark_PN}\cite{sign:sign
diquark_LamdaO}\cite{sign:sign diquark_Lamda}, and some baryons
can be rewritten as\cite{sign:sign diquark_baryon}
\begin{eqnarray}
 \nonumber
 \mid{\Lambda}\rangle &=&\frac{1}{\sqrt{3}}[B_\frac{1}{2}(S_{ud},s)+\sqrt{\frac{3}{4}}B_\frac{1}{2}(V_{us},d)-\sqrt{\frac{3}{4}}B_\frac{1}{2}(V_{ds},u)+\sqrt{\frac{1}{4}}B_\frac{1}{2}(S_{us},d)-\sqrt{\frac{1}{4}}B_\frac{1}{2}(S_{ds},u)],\\
 \nonumber
 \mid{\Sigma^0}\rangle
 &=&\frac{1}{\sqrt{3}}[B_\frac{1}{2}(V_{ud},s)-\sqrt{\frac{1}{4}}B_\frac{1}{2}(V_{us},d)-\sqrt{\frac{1}{4}}B_\frac{1}{2}(V_{ds},u)+\sqrt{\frac{3}{4}}B_\frac{1}{2}(S_{us},d)+\sqrt{\frac{3}{4}}B_\frac{1}{2}(S_{ds},u)],\\
\nonumber
 \mid{\Omega^{-}}\rangle &=&B_\frac{3}{2}(V_{ss},s),
\end{eqnarray}
where B represent a baryon state.

So the productions can be described as
\begin{eqnarray}
\nonumber
 \frac{d \Lambda}{dt}&=&\frac{1}{3}\cdot[\frac{3}{4}\Gamma(V_{us},d,\Lambda)+\frac{3}{4}\Gamma(V_{ds},u,\Lambda)
 +\frac{1}{4}\Gamma(S_{us},d,\Lambda)+\frac{1}{4}\Gamma(S_{ds},u,\Lambda)+\Gamma(S_{ud},s,\Lambda)],\\
\nonumber
 \frac{d
 \Sigma^0}{dt}&=&\frac{1}{3}\cdot[\Gamma(V_{ud},s,\Sigma^0)+\frac{1}{4}\Gamma(V_{us},d,\Sigma^0)+\frac{1}{4}\Gamma(V_{ds},u,\Sigma^0)+\frac{3}{4}\Gamma(S_{us},d,\Sigma^0)+\frac{3}{4}\Gamma(S_{ds},u,\Sigma^0)],\\
 \frac{d \Omega^{-}}{dt}&=&\Gamma(V_{ss},s,\Omega^{-}),
\end{eqnarray}
$p$, $n$, $\Xi^0$, $\Xi^-$ and other baryons can be calculated
through similar methods.

As a simplification, baryon production can be described as a
combination of different processes of quarks and diquarks forming
$(\frac{1}{2})^+$ or $(\frac{3}{2})^+$ baryon states, as
\begin{equation}
 \frac{dB}{dt}=\sum{C_{cg}^2(D_{q_1q_2},q_3,B)\Gamma(D_{q_1q_2},q_3,B)},
\end{equation}
where $C_{cg}^2(D_{q_1q_2},q_3,B)$ is the Clebsch-Gordan
coefficient to represent the state of quark-diquark coupling shown
in equations (1), and one could get the result\cite{sign:sign
diquark_baryon}\cite{sign:sign crossection} after the integration
of the producing cross-sections of the baryon states above under
the conditions of quarks and diquarks are assumed to be under
ideal Fermi and Bose distributions.
\begin{equation}
 \frac{dB}{dt}=\sum{C_{cg}^2(D_{q_1q_2},q_3,B)\frac{3\omega_D\omega_q|M|^2}{32\pi^2}T^2F_{FB}(q_3,D_{q_1q_2},B,T)},
\end{equation}
where $\omega_D$ and $\omega_q$ are the spin and color degeneracy
of quarks and diquarks, while
\begin{eqnarray*}
 T^2F_{FB}(q,D,B,T)=\int\int\frac{dE_qdE_{D}}{(e^{\frac{E_q-\mu_q}{T}}+1)(e^{\frac{E_{D}-\mu_{D}}{T}}-1)},
\end{eqnarray*}
The integrating ranges are
\begin{eqnarray*}
 m_q\leq{E_q}\leq{\infty},
 m_{D}\leq{E_{D}}\leq{\infty}
\end{eqnarray*}
and

\[E_qE_{D}\geq{\frac{1}{4m_B^2}\{4(E_q+E_{D})(m_q^2E_{D}+m_{D}^2E_q)+[m_B^2-(m_q+m_{D})^2][m_B^2-(m_q-m_{D})^2]\}}
\]

For $(\frac{1}{2})^+$ baryons from axial-vector diquarks, one has
the effective lagrangian (with some corrections of the
expressions in\cite{sign:sign diquark_baryon}, which make few
differences on the final results.)
\begin{equation}
 L_{intV_{1/2}} = ig\bar{B}\gamma_{\mu}\gamma_{5}q V_{\mu},
\end{equation}
Then,
\begin{equation}
 |M|^2_{V_{1/2}} =\frac{g^2}{3}[\frac{(m_B^2-m_q^2)^2}{m_{V}^2}+m_B^2+m_q^2-2m_{V}^2+6m_Bm_q],
\end{equation}

For $(\frac{1}{2})^+$ baryons from  scalar diquarks,

\begin{equation}
 L_{intS_{1/2}} = ig\bar{B}q S,
\end{equation}
\begin{equation}
 |M|^2_{S_{1/2}} =
 {g^2}[(m_B+m_q)^2-m_{S}^2],
\end{equation}

For $(\frac{3}{2})^+$ baryons, the $|M|^2$ are rather complicated
and only ratios of anti-baryon/baryon are calculated, as the same
matrices.

The diquark mass is preliminarily assumed as
$m_D(q_1q_2)=m_{d0}+m_{q_1}+m_{q_2}$ and  the difference of
axial-vector diquark mass and scalar diquark mass is neglected as
a simple assumption. $m_{d0}$ here is about 400-800 $MeV$ and
should not be smaller than the masses of constituent quarks.
Additionally, it is assumed that $\bf{g}$ is same in these
reactions.

\vspace{12pt}

\leftline{ \bf 3. Analysis}

\vspace{8pt}

In the calculations, we set current quark mass,
\[ m_u = 3 MeV,    m_d = 6 MeV,     m_s = 122.5 MeV, \]
as the mean masses. Different quark masses may cause systematic
errors about $5\%\sim10\%$. The critical temperature is estimated
at $T_c{_{(\mu=0)}}=166.1MeV$ from a recent calculation based on
\cite{sign:sign Strange} and \cite{sign:sign diOmega}. The
critical temperatures estimated from different methods (such as
\cite{sign:sign TC_lattice1}\cite{sign:sign TC_lattice2},) are
similar and may cause systematic errors about $1\%\sim5\%$.

$\overline{\Omega}^+/\Omega^-$ ratios calculated in different
conditions are shown in Figure 1. It is clearly that the ratio is
always greater than 1 when $m_{d0}$ and $\mu_B$ varies, even if
ideal Fermi or Bose distributions are not formed. (*P.S. RHIC
data\cite{sign:sign Omega/OmegaRHIC_1} of
$\overline{\Omega}^+/\Omega^-$ at $130$ $GeV$ is before
corrections such as annihilation of the daughter anti-protons with
physical material in the detectors. After that correction the
ratio may larger than 1, see in \cite{sign:sign pp_STAR_Omega} Fig
6.)

Anti-baryon/baryon ratios are listed in Table 1, $\mu_B \approx
47$ $MeV$ and $\mu_B \approx 30$ $MeV$ are used to meet data from
RHIC Au-Au at $\sqrt{s_{NN}} = 130$ $GeV$ and $200$
$GeV$\cite{sign:sign Omega/OmegaRHIC_1}\cite{sign:sign
pp_STAR_Omega}\cite{sign:sign xunu_pp061}\cite{sign:sign
theo_fit}\cite{sign:sign harris}\cite{sign:sign
pp_PHOBOS_073}\cite{sign:sign pp_PHENIX_070}\cite{sign:sign
PHENIX_sum}\cite{sign:sign pp_Star130_065}\cite{sign:sign
Lambda_STAR}\cite{sign:sign Lambda_PHENIX}\cite{sign:sign
STAR_Lambda_200}, as $m_{d0}$ is estimated at 600 $MeV$. Many of
the calculations work well except $\Xi$ productions, which also
promote the inclusive $\Lambda$ productions. Some strange baryon
over proton ratios are listed in Table 2 and 3, compared with the
upper limits from the Hadronic Gas Model. Some theoretical values
are larger than the PHENIX preliminary results\cite{sign:sign
Lambda_PHENIX}(, for which HG model works well). This may be
caused by the condition that ideal QGP fluid is not completely
formed at the temperature of $T_c$.

\begin{figure}[htb]
  \centering
  \includegraphics[width=0.7\textwidth]{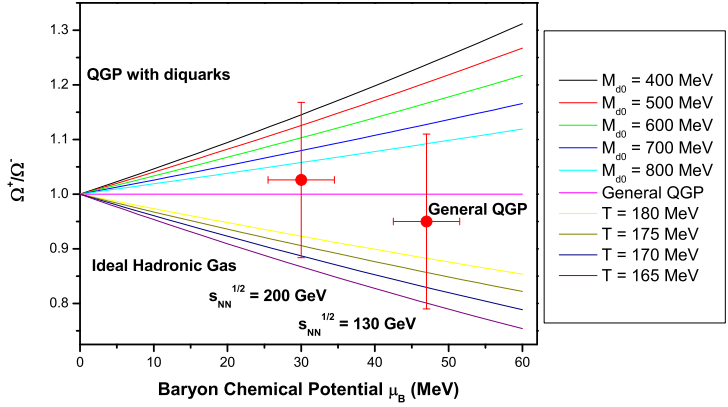}
  \caption{\quad Ratios of $\overline{\Omega}^+/\Omega^-$ from QGP with
diquarks (lines larger than 1.0), general QGP(line at 1.0) and the
upper limits of Hadronic Gas Model (lines smaller than 1.0) at
different conditions, compared with preliminary results from
RHIC\cite{sign:sign Omega/OmegaRHIC_1}. Systematic errors $\sim
5\%$.}
\end{figure}

\begin{center}
\begin{tabular}{| l | c | c | c | c |}
 \hline & $\sqrt{s_{NN}} = 130$ $GeV$ & $\mu_B = 47$ $MeV$ & $\sqrt{s_{NN}} =200$ $GeV$ & $\mu_B = 30$ $MeV$ \\
 \hline $Ratios$  & $Exp.$ & $QGP_d$ &  $Exp.$ & $QGP_d$  \\
 \hline $\bar{p}/p$                                                      & -                    & $0.579$ & - & $0.706$ \\
 \hline $\bar{p}/p$ $_{(p,}$ $_{\Sigma^+)}$                              & $0.61 \pm 0.03 \pm 0.06$ \cite{sign:sign harris}& $0.616$ & $0.73 \pm 0.02 \pm 0.03$ \cite{sign:sign pp_PHOBOS_073} & $0.734$ \\
                                                                         &&&$0.70 \pm 0.04 \pm0.10 $ \cite{sign:sign pp_PHENIX_070}&  \\
                                                                         &&&$0.731\pm 0.011\pm0.062$ \cite{sign:sign PHENIX_sum}&  \\
 \hline $\bar{p}/p$$_{(inclusive)}$                                      & $0.72 \pm 0.05$ \cite{sign:sign pp_STAR_Omega} & $0.698$ & $0.78 \pm 0.05$ \cite{sign:sign pp_STAR_Omega}         & $0.795$ \\
                                                                         & $0.65 \pm 0.01 \pm 0.07$ \cite{sign:sign pp_Star130_065}&&$0.747\pm0.007\pm0.046$ \cite{sign:sign PHENIX_sum}&  \\
 \hline $\frac{\overline{\Lambda}+\overline{\Sigma}^0}{\Lambda+\Sigma^0}$& $0.73 \pm 0.03$ \cite{sign:sign harris}& $0.726$ & - & $0.815$\\
 \hline $\frac{\overline{\Lambda}}{\Lambda}$ $_{(inclusive)}$            & $0.74 \pm 0.04 \pm 0.03$ \cite{sign:sign Lambda_STAR}& $0.798$ & $\approx>0.8$ \cite{sign:sign STAR_Lambda_200}& $0.866$\\
                                                                         & $0.75 \pm 0.09 \pm 0.17$ \cite{sign:sign Lambda_PHENIX}&&&  \\
 \hline $\frac{\overline{\Xi}^0+\overline{\Xi}^+}{\Xi^0+\Xi^-}$          & $0.82 \pm 0.08$ \cite{sign:sign harris}& $0.921$ & - & $0.949$\\
 \hline $\overline{\Omega}^+/\Omega^-$                                   & $0.95 \pm 0.15 \pm0.05^*$ \cite{sign:sign Omega/OmegaRHIC_1}& $1.166$ & $1.026 \pm 0.075\pm0.12$ \cite{sign:sign Omega/OmegaRHIC_1}& $1.103$ \\
 \hline
\end{tabular}
\end{center}
{Table 1: \quad Anti-baryon/baryon ratios at $m_{d0}$ = 600 $MeV$.
Where $p$ $_{(p,}$ $_{\Sigma^+)}$ includes the decay
contributions of $\Sigma^+$, $p$ $_{(inclusive)}$ includes the
decay contributions of $\Sigma^+$ and $\Lambda$ baryons,
$\Lambda$ $_{(inclusive)}$ includes the decay contributions of
$\Xi^0$ and $\Xi^-$, contributions of $\Omega$ are neglected.
Systematic errors $\sim 5\%$.}

\begin{center}
\begin{tabular}{|l|c|c|c|}

 \hline $Ratios$ & $Exp.$ & $QGP_d$ & $HG$ $limit$ \\
 \hline $\frac{\Lambda+\Sigma^0}{p}$                                                  & - & $1.205$ & $0.695$  \\
 \hline $\frac{\overline{\Lambda}+\overline{\Sigma}^0}{\bar{p}}$                      & - & $1.510$ & $0.785$  \\
 \hline $\frac{\Lambda(inclusive)}{p(p, \Sigma^+)}$                                   & $0.89 \pm 0.07 \pm 0.21$ \cite{sign:sign Lambda_PHENIX}& $1.282$ & $0.858$  \\
 \hline $\frac{\overline{\Lambda}(inclusive)}{\bar{p}(\bar{p},\overline{\Sigma}^-)}$  & $0.95 \pm 0.09 \pm 0.22$ \cite{sign:sign Lambda_PHENIX}& $1.659$ & $0.989$  \\
 \hline $\frac{\Sigma^0}{\Lambda}$                                                    & - & $1.307$ & $0.692$  \\
 \hline $\frac{\overline{\Sigma}^0}{\overline{\Lambda}}$                              & - & $1.291$ & $0.692$  \\
 \hline
\end{tabular}
\end{center}
{Table 2 \quad Relative yields of baryons at $\sqrt{s_{NN}} = 130
GeV$ and $\mu_B = 47 MeV$ from QGP with diquarks compared with
Ideal Hadronic Gas limit at $T=170$ $MeV$. Systematic errors $\sim
15\%$.}

\begin{center}
\begin{tabular}{|l|c|c|c|}

 \hline $Ratios$ & $Exp.$ & $QGP_d$ & $HG$ $limit$ \\
 \hline $\frac{\Lambda+\Sigma^0}{p}$                                                  & - & $1.256$ & $0.710$  \\
 \hline $\frac{\overline{\Lambda}+\overline{\Sigma}^0}{\bar{p}}$                      & - & $1.451$ & $0.767$  \\
 \hline $\frac{\Lambda(inclusive)}{p(p, \Sigma^+)}$                                   & - & $1.347$ & $0.880$  \\
 \hline $\frac{\overline{\Lambda}(inclusive)}{\bar{p}(\bar{p},\overline{\Sigma}^-)}$  & - & $1.588$ & $0.964$  \\
 \hline $\frac{\Sigma^0}{\Lambda}$                                                    & - & $1.305$ & $0.692$  \\
 \hline $\frac{\overline{\Sigma}^0}{\overline{\Lambda}}$                              & - & $1.294$ & $0.692$  \\
 \hline
\end{tabular}
\end{center}
{Table 3 \quad Relative yields of baryons at $\sqrt{s_{NN}} = 200
GeV$ and $\mu_B = 30 MeV$  from QGP with diquarks compared with
Ideal Hadronic Gas limit at $T=170$ $MeV$. Systematic errors
$\sim 15\%$.}

\vspace{4pt}

\vspace{12pt}

\leftline{ \bf 4. Discussion}

\vspace{8pt}

Baryon production ratios could be researched in the model of
Quark-Gluon Plasma with diquarks and strange baryon over proton
ratios from these calculations are larger than those of Hadronic
Gas\cite{sign:sign diquark_baryon}. Another result is that
$\Sigma^0$ production is larger than $\Lambda$, which in Hadronic
Gas Model is smaller. But it is hard to be observed, due to the
short decay length. $\overline{\Omega}^+/\Omega^- > 1$ is the
most unusual results, which has been slightly supported by
$\Omega^-/h^-$ and $\overline{\Omega}^+/h^-$ measured in
RHIC\cite{sign:sign Omega/OmegaRHIC_1}, although the statistical
and systematic errors are too large to confirm it.

It is reported that local thermal equilibrium has been formed in
RHIC with high energy density based on recent hydrodynamic
analysis\cite{sign:sign Hydro}. As well as other evidence such as
jet quenching\cite{sign:sign JET}, QGP existing in RHIC is nearly
proved. $\overline{\Omega}^+/\Omega^-
> 1$ will be another strong evidence if future results with
smaller errors could confirm it.

\vspace{12pt}

\leftline{ \bf Acknowledgement }

\vspace{8pt}

We would like to thank Doctor Gene Van Buren and Doctor
Christophe Suire for helpful discussions about experimental data.
This work was supported in part by the National Natural Science
Foundation of China (90103019), and the Doctoral Programme
Foundation of Institution of Higher Education, the State
Education Commission of China (2000000147).



\begin{thebibliography}{99} \small
\bibitem[1]{sign:sign diquark_form} Pavkovi$\acute{c}$ M I, Phys. Rev. D. {\bf13} (1976) 2128.

\bibitem[2]{sign:sign bound_state}  Shuryak E, Zahed I, hep-ph/0403127.

\bibitem[3]{sign:sign screen}Mustafa M, Thoma M, Chakraborty P, hep-ph/0403279.


\bibitem[4]{sign:sign diquark_baryon} Miao H, Ma Z B and Gao C S, J. Phys. {\bf{G29}} (2003) 2187-2192.

\bibitem[5]{sign:sign Rafelski_phL} Rafelski J, Letessier J, Phys. Lett {\bf B469}
(1999) 12. nucl-th/9908024.

\bibitem[6]{sign:sign Rafelski_Acta1} Rafelski J, Letessier J, Tounsi A, Acta Phys. Polon. {\bf B27} (1996)
1037.

\bibitem[7]{sign:sign Rafelski_Acta2} Rafelski J, Letessier J, Acta Phys. Polon. {\bf B30} (1999)
3559, hep-ph/9910300.

\bibitem[8]{sign:sign Rafelski_Acta3} Letessier J, Rafelski J, Acta Phys. Polon. {\bf B30} (1999)
153.

\bibitem[9]{sign:sign Strange} Gao C S, Wu T, J. Phys. {\bf{G27}} (2001) 459-463.

\bibitem[10]{sign:sign RatiosModel} Braun-Munzinger P, Redlich K, Stachel J, Invited review for Quark Gluon Plasma 3,
 eds. Hwa R C and Wang X N, World Scientific Publishing, nucl-th/0304013.

\bibitem[11]{sign:sign diquark_density} Ma Z B, Miao H and Gao C S, Chin. Phys. Lett. {\bf{20}} (2003) 1691-1693.

\bibitem[12]{sign:sign Omega/OmegaSPS} Mitrovski M, ( for the NA49 collaboration), J. Phys. {\bf{G30}} (2004) S357-S362.

\bibitem[13]{sign:sign Omega/OmegaString} Bleicher M, et al., Phys. Rev. Lett. {\bf{88}} (2002) 202501.

\bibitem[14]{sign:sign diquark_PN} Ma B Q, Qing D, Schmidt I, Phys. Rev. {\bf{C65}} (2002) 035205, hep-ph/0202015.

\bibitem[15]{sign:sign diquark_LamdaO} Carlitz R, Phys. Lett. {\bf{B58}} (1975) 345; Kaur J, Nucl. Phys. {\bf{B128}} (1977) 219; Sch$\ddot{a}$fer A, Phys. Lett. {\bf{B208}} (1988) 175.

\bibitem[16]{sign:sign diquark_Lamda} Ma B Q, Schmidt I, Yang J J, Phys. Lett. {\bf{B477}} (2000) 107-113, hep-ph/9906424.

\bibitem[17]{sign:sign crossection} Gao C S, in JingShin Theoretical Physics Symposium in Honor of Professor Ta-You Wu T Y, edited by Hsu J P and Hsu L, (World Scientific, 1998) 362.

\bibitem[18]{sign:sign diOmega} Gao C S, Commun. Theor. Phys. {\bf{40}} (2003)
188.
\bibitem[19]{sign:sign TC_lattice1} Karsch F, Nucl. Phys. (Proc Suppl) 83-84 (2000) 14.

\bibitem[20]{sign:sign TC_lattice2} Fodor Z, Katz S, JHEP 0203 (2002) 014, hep-lat/0106002.

\bibitem[21]{sign:sign Omega/OmegaRHIC_1} Suire C, (for the STAR Collaboration), Nucl. Phys. {\bf{A715}} (2003)
470-4735, nucl-ex/0211017; Suire C, (for the STAR Collaboration),
Quark Matter 2002.

\bibitem[22]{sign:sign pp_STAR_Omega} Van Buren G, (for the STAR Collaboration), Nucl. Phys. {\bf{A715}} (2003)
129-139, nucl-ex/0211021.

\bibitem[23]{sign:sign xunu_pp061} Xu N, Kaneta M, Nucl. Phys. {\bf{A698}} (2002) 306-313, nucl-ex/0104021.

\bibitem[24]{sign:sign theo_fit} Baran A, Broniowski W and Florkowski
W, Acta Phys. Polon. {\bf{B35}} (2004) 779-798, nucl-th/0305075.

\bibitem[25]{sign:sign harris} Harris J, Overview of First Results from Star, (for STAR Collaboration), Quark Matter 2002.

\bibitem[26]{sign:sign pp_PHOBOS_073} Wosiek B, (for the PHOBOS Collaboration), Nucl. Phys. {\bf{A715}} (2003)
510-513, nucl-ex/0210037.

\bibitem[27]{sign:sign pp_PHENIX_070} Chujo T, (for the PHENIX
Collaboration), Nucl. Phys. {\bf{A715}} (2003) 151-160,
nucl-ex/0209027.

\bibitem[28]{sign:sign PHENIX_sum} Adler S S, et al., PHENIX
Collaboration, nucl-ex/0307022.

\bibitem[29]{sign:sign pp_Star130_065} Adler C, et al., the STAR Collaboration, Phys. Rev. Lett. {\bf{86}} (2001) 4778-4782.

\bibitem[30]{sign:sign Lambda_STAR} Adler C, et al., the STAR Collaboration, Phys. Rev. Lett. {\bf{89}} (2002) 092301.

\bibitem[31]{sign:sign Lambda_PHENIX} Adcox K, et al., the PHENIX
Collaboration, Phys. Rev. Lett. {\bf{89}} (2002) 092302.

\bibitem[32]{sign:sign STAR_Lambda_200} Van Buren G, (for the STAR Collaboration), J. Phys. {\bf{G28}} (2002) 2103-2108, nucl-ex/0201009.

\bibitem[33]{sign:sign Hydro} Hirano T, Quark Matter 2004.
\bibitem[34]{sign:sign JET} Back B B, et al., the PHOBOS
Collaboration, Phys. Rev. Lett. {\bf{91}} (2002) 072302; Adler S
S, et al., the PHENIX Collaboration, Phys. Rev. Lett. {\bf{91}}
(2002) 072303; Adams J, et al., the STAR Collaboration, Phys.
Rev. Lett. {\bf{91}} (2002) 072304; Arsene I, et al., the BRAHMS
Collaboration, Phys. Rev. Lett. {\bf{91}} (2002) 072305.




\end{thebibliography}
\end{document}